\documentclass[conference, a4paper]{IEEEtran}
\IEEEoverridecommandlockouts
\def\BibTeX{{\rm B\kern-.05em{\sc i\kern-.025em b}\kern-.08em
    T\kern-.1667em\lower.7ex\hbox{E}\kern-.125emX}}

\usepackage{amsmath,amssymb,bm,cite,algorithm,algorithmicx,amsthm}
\usepackage{algpseudocode,fancyhdr,amsfonts,cases,graphicx,xcolor}
\usepackage{multirow,mathtools,tabularx,epsfig}
\usepackage[english]{babel}
\usepackage[top=.75in, left= 0.625in, right = 0.63in, bottom =1in]{geometry}

\newtheorem{proposition}{Proposition}
\DeclareMathOperator{\sinc}{sinc}
\IEEEoverridecommandlockouts

\begin{document}
\title{Uplink Beam Management for Millimeter Wave Cellular MIMO Systems with Hybrid Beamforming}
\author{
\IEEEauthorblockN{George~C.~Alexandropoulos$^1$, Ioanna Vinieratou$^1$, Mattia Rebato$^{2}$, Luca Rose$^{3}$, and Michele Zorzi$^{2}$
}
\IEEEauthorblockA{
$^1$Department of Informatics and Telecommunications, National and Kapodistrian University of Athens, Greece\\
$^2$Department of Information Engineering, University of Padova, 35131 Padova, Italy\\
$^3$Nokia Bell Labs, 91620 Nozay, France\\
emails: \{alexandg, en2180001\}@di.uoa.gr, \{rebatoma, zorzi\}@dei.unipd.it, luca.rose@nokia.com}}

\maketitle

\begin{abstract}
Hybrid analog and digital BeamForming (HBF) is one of the enabling transceiver technologies for millimeter Wave (mmWave) Multiple Input Multiple Output (MIMO) systems. This technology offers highly directional communication, which is able to confront the intrinsic characteristics of mmWave signal propagation. However, the small coherence time in mmWave systems, especially under mobility conditions, renders efficient Beam Management (BM) in standalone mmWave communication a very difficult task. In this paper, we consider HBF transceivers with planar antenna panels and design a multi-level beam codebook for the analog beamformer comprising flat top beams with variable widths. These beams exhibit an almost constant array gain for the whole desired angle width, thereby facilitating efficient hierarchical BM. Focusing on the uplink communication, we present a novel beam training algorithm with dynamic beam ordering, which is suitable for the stringent latency requirements of the latest mmWave standard discussions. Our simulation results showcase the latency performance improvement and received signal-to-noise ratio with different variations of the proposed scheme over the optimum beam training scheme based on exhaustive narrow beam search.
\end{abstract}


\section{Introduction}\label{sec:Intro}
Mobile communication in the millimeter Wave (mmWave) \cite{Rappaport_Access_all} and THz \cite{Akyildiz_COMmag2015} frequency bands is a promising technology for the increasingly demanding data rate requirements of beyond fifth Generation (5G) wireless networks. Short-range mmWave communication (up to $10\,{\rm m}$) in the unlicensed band around $60\,{\rm GHz}$ with a maximum of $2.16\,{\rm GHz}$ bandwidth is standardized in IEEE 802.11ad \cite{J:IEEE_11ad} offering up to $7\,{\rm GBps}$ data rate. The follow-up IEEE 802.11ay standard (to be finalized till the end of 2020 for up to $500\,{\rm m}$ transmission distance), which shares physical layer features with the 5G New Radio (NR) technology for the mmWave band \cite{J:5G_NR_2017}, is intended to have a maximum bandwidth of $8.64\,{\rm GHz}$ increasing the peak data rate to $100\,{\rm GBps}$. Among the technical enablers for this extensive rate improvement compared to IEEE 802.11ad are the channel bonding, higher order modulation, multiple independent data streams transmissions, and multi-user Multiple Input Multiple Output (MIMO) capabilities.   
  
Transmitted signals at mmWave communications experience severe path and penetration losses as well as rain fading. However, their mm-level wavelengths enable many antenna elements to be packed in small physical dimensions, which allows for highly directional BeamForming (BF). IEEE 802.11ad adopts analog BF \cite{Venkateswaran_ABF} with single-stream communication, where many antenna elements are connected to one Radio Frequency (RF) chain via an analog preprocessing network (usually comprised of phase shifters). Note that connecting each antenna element to a dedicated RF chain, as in typical MIMO systems with carrier frequencies below $6$GHz, is prohibitive at mmWave frequencies with up to date radios (e$.$g$.$, see \cite{J:Doan_CMOS}) from both cost and power consumption perspectives. To provide higher throughputs, the IEEE 802.11ay standard will support Hybrid BF (HBF) \cite{J:Molisch_HBF, Molisch_HBF_2017_all} with up to four spatial communication streams. This technology, that capitalizes on both analog and digital processing, enables a large number of antennas to be connected (all or in groups) to a much smaller number of RF chains. The latter number determines the maximum number of independent data streams that can be communicated concurrently, whereas the BF gain depends on the total number of antenna elements. 

As a channel-dependent process, BF requires Channel State Information (CSI) knowledge availability at the communication end that applies it. However, the RF hardware limitations with HBF architectures in conjunction with the short coherence time in mmWave systems, which becomes smaller under mobility conditions, renders CSI very hard to acquire. Although recent works (e.g., \cite{Alkhateeb_JSTSP_all, Marzi_TSP2016_all, Bazzi_SPAWC2016_all, Alexandropoulos_GC16, Vlachos_SPL18, Vlachos2019WidebandSampling}) capitalized on the spatial sparsity of mmWave channels \cite{Zhang_WCNC2010_all} to estimate portions or the entire MIMO channel matrix, the presented approaches required lengthy beam training phases. Those approaches aimed at estimating the channel in multiple directions using machine learning tools. Another family of techniques for efficient analog BF (e$.$g$.$, \cite{Hur_TCOM_all, Hosoya_TAP2015_all, Wang_JSAC_all, Jeong_COMmag2015, Barati_2015, Raghavan_JSTSP, Alexandropoulos_position, Aykin_2020}) is based on beam switching between the communicating nodes in order to find a pair of beams from their available beam codebooks meeting a link performance indicator threshold. When such a beam pair is found, beam alignment is considered achieved and no further beam searching is needed. An iterative algorithm for HBF design with practical beam codebooks has been also presented in \cite{Love_HBF_Codebook_2018}.

In this paper, we focus on the 5G NR Beam Management (BM) specifications \cite{J:5G_NR_2017} and present a beam training algorithm for the UpLink (UL) of mmWave MIMO communication with HBF transceivers. The presented algorithm capitalizes on our novel multi-level beam codebook for analog BF with planar antenna panels, which comprises Flat Top Beams (FTBs) with variable widths. The designed beams exhibit an almost constant array gain across their entire beamwidth facilitating efficient hierarchical beam sweeping for initial access and beam alignment recovery, as well as for data communication. Our simulation results obtained with the realistic mmWave channel model \cite{akdeniz14_all}, demonstrate the latency improvement offered by the proposed BM algorithm over benchmarked variations of narrow beam searching for mmWave UL communication.  

\textit{Notation:} Vectors and matrices are denoted by boldface lowercase and boldface capital letters, respectively. $\mathbf{A}$'s Hermitian transpose is denoted $\mathbf{A}^{\rm H}$, ${\rm vec}(\mathbf{A})$ represents a column vector resulting from stacking $\mathbf{A}$'s columns one below another, and $\mathbf{I}_{n}$ ($n\geq2$) is the $n\times n$ identity matrix. 
$\mathbf{0}_{m\times n}$ ($\mathbf{0}_{m}$) is the $m\times n$ ($m\times 1$) matrix of all zeros, while $\mathbf{1}_{m\times n}$ is the all ones $m\times n$ matrix and $[\boldsymbol{\alpha}]_i$ stands for $\boldsymbol{\alpha}$'s $i$-th element. $\mathbb{R}$ and $\mathbb{C}$ represent the real and complex number sets, respectively, ${\rm card}(\cdot)$ denotes set cardinality, and $x\sim\mathcal{C}\mathcal{N}\left(0,\sigma^{2}\right)$ indicates that $x$ is a circularly symmetric complex Gaussian random variable with zero mean and variance $\sigma^{2}$. 

\vspace{-0.5mm}
\section{System and Channel Models}\label{sec:System_Model}
In this section, we present the system model under investigation as well as the considered wireless channel model suitable for mmWave cellular MIMO communication with HBF transceivers.

\vspace{-0.5mm}
\subsection{System Model}
We consider a mmWave point-to-point communication system comprising one User Equipment (UE) having $N_{\rm ue}$ antennas and a Base Station (BS) equipped with $N_{\rm bs}$ antennas. The antenna elements at the UE are assumed to form a linear array, while the antenna elements at the BS are placed on planar panels. More specifically, the coplanar BS antennas are grouped into $M_{\rm bs}\ll N_{\rm bs}$ planar antenna packages, termed hereinafter as Antennas in Packages (AiPs) \cite{AiP_2009}. Each of these AiPs is comprised of one $N_{\rm bs,h}\times N_{\rm bs,v}$ planar array, where $N_{\rm bs,h}$ and $N_{\rm bs,v}$ denote the number of antenna elements in the horizontal and vertical directions, respectively; it clearly holds that $M_{\rm bs}N_{\rm bs,h}N_{\rm bs,v}=N_{\rm bs}$. The number of RF chains at the UE is assumed to be $N_{\rm ue}$ with each UE antenna connected to a distinct RF chain, and each AiP at the BS side is considered connected to a dedicated RF chain. Hence, $M_{\rm bs}$ RF chains in total are available at the BS, which possesses HBF capability, while the UE performs fully digital BF. 

Among the available HBF architectures \cite{Molisch_HBF_2017_all}, belongs to the AiP-based (or partially connected) architecture, according to which the analog beamformer has a specific block diagonal structure. In this paper, we focus on UL communication and assume that the BS employs the following $N_{\rm bs}\times M_{\rm bs}$ analog combiner:
\begin{equation}\label{EQ:PC_Analog_Combiner}
\mathbf{U}_{\rm RF} = \left[ \begin{matrix}
\mathbf{u}_1  &  \mathbf{0}_{N_{\rm bs,h}N_{\rm bs,v}}    &  \cdots    &  \mathbf{0}_{N_{\rm bs,h}N_{\rm bs,v}}  \\
\mathbf{0}_{N_{\rm bs,h}N_{\rm bs,v}}    &  \mathbf{u}_2  &  \cdots    &  \mathbf{0}_{N_{\rm bs,h}N_{\rm bs,v}}  \\
\vdots   &  \vdots   &  \ddots    &	 \vdots  \\
\mathbf{0}_{N_{\rm bs,h}N_{\rm bs,v}}    &  \mathbf{0}_{N_{\rm bs,h}N_{\rm bs,v}}    &  \cdots    &  \mathbf{u}_{M_{\rm bs}} 
\end{matrix}
\right],
\end{equation}
where the complex-valued $N_{\rm bs,h}N_{\rm bs,v}$-element column vector $\mathbf{u}_i$ with $i=1,2,\dots,M_{\rm bs}$ represents the BF vector at the BS's $i$-th AiP, whose elements are assumed to have unit magnitude, i$.$e$.$, $|[\mathbf{u}_i]_{n}|^2=1$ $\forall$$n=1,2,\ldots,N_{\rm bs,h}N_{\rm bs,v}$. In addition, we assume that $\mathbf{u}_i\in\mathbb{F}_{{\rm bs}}$ $\forall$$i$, which means that the BF vectors at all BS's AiPs belong to a predefined beam codebook $\mathbb{F}_{{\rm bs}}$ including ${\rm card}(\mathbb{F}_{\rm bs})$ distinct vectors (beams). The $M_{\rm bs}$ outputs of the BS's AiPs are finally connected to this node's digital processing block whose operation is described by the digital BF $\mathbf{U}_{\rm BB}\in\mathbb{C}^{M_{\rm bs}\times N_s}$, where $N_s\leq\min(N_{\rm ue},M_{\rm bs})$ determines the number of data streams that can be concurrently communicated through the $N_{\rm bs}\times N_{\rm ue}$ UL MIMO channel, whose gains are included in $\mathbf{H}\in\mathbb{C}^{N_{\rm bs}\times N_{\rm ue}}$. The latter matrix is obtained by the matrix row concatenation $\mathbf{H}\triangleq[\mathbf{H}_1;\mathbf{H}_2;\cdots;\mathbf{H}_{M_{\rm bs}}]$, where $\mathbf{H}_i\in\mathbb{C}^{N_{\rm bs,h}N_{\rm bs,v}\times N_{\rm ue}}$ denotes the UL MIMO channel gain matrix between BS's $i$-th AiP and the UE. We finally consider digital precoding of the transmitted signals represented by $\mathbf{V}_{\rm BB}\in\mathbb{C}^{N_{\rm ue}\times N_s}$ at the UE side. 

The considered mmWave UL communication is realized on a block-by-block basis, where each block comprises of two distinct phases: \textit{i}) a control phase for sounding the channel and establishing data communication, and \textit{ii}) a data phase for communicating UE's signal. The control phase further constitutes of distinct Control Time Slots (CTSs) during which UE's and BS's HBF matrices are jointly designed according to certain performance objectives. Following the latter definitions, the BaseBand (BB) received signal at the $M_{\rm bs}$ outputs of the BS AiPs can be mathematically expressed via the following expression:
\begin{equation}\label{Eq:UL_Received_Signal_UE_BB}
\mathbf{y} = \mathbf{U}_{\rm RF}^{\rm H}\mathbf{H}\mathbf{x} + \mathbf{U}_{\rm RF}^{\rm H}\mathbf{n},
\end{equation}
where the complex-valued $N_{\rm ue}$-element column vector $\mathbf{x}$ contains the UE's precoded information data or training symbols and $\mathbf{n}\in\mathbb{C}^{N_{\rm bs}\times 1}$ represents the zero-mean Additive White Gaussian Noise (AWGN) vector with covariance matrix $\sigma^2\mathbf{I}_{N_{\rm bs}}$. During the data communication phase, vector $\mathbf{x}$ is expressed as $\mathbf{x}\triangleq\mathbf{V}_{\rm BB}\mathbf{s}$ with the complex-valued $N_s$-element column vector $\mathbf{s}$ including the unit power data symbols chosen from a discrete modulation set. In the channel sounding phase, training symbols known at both communications ends are transmitted in the place of data. Before data detection or channel estimation at the BS side, the received UE symbols in \eqref{Eq:UL_Received_Signal_UE_BB} are processed as $\mathbf{U}_{\rm BB}^{\rm H}\mathbf{y}$ using the digital BF matrix $\mathbf{U}_{\rm BB}\in\mathbb{C}^{N_{\rm bs}\times N_s}$.
\vspace{-0.5mm}
\subsection{Channel Model}\label{Sec:Channel_Model}
In this paper, we adopt the NYU mmWave channel model \cite{akdeniz14_all} for the $N_{\rm bs}\times N_{\rm ue}$ UL MIMO matrix $\mathbf{H}$, which is based on extensive real-world measurement campaigns in New York City in the $28\,{\rm GHz}$ frequency band. This model is actually based on the WINNER II model \cite{winner2} and considers that each wireless link comprises $K$ clusters, corresponding to macro-level scattering paths, each composed of $L$ sub-paths. To this end, the mmWave MIMO channel $\mathbf{H}$ is modeled at an arbitrary coherent time instant as \cite[eq. (9)]{akdeniz14_all}
\begin{equation}
\mathbf{H}= \sum_{k=1}^{K}\sum_{\ell=1}^{L}g_{k,\ell} \mathbf{a}_{\rm bs}\left(\theta^{\rm bs}_{k,\ell},\phi^{\rm bs}_{k,\ell}\right) \mathbf{a}^{\rm H}_{\rm ue}\left(\theta^{\rm ue}_{k,\ell},\phi^{\rm ue}_{k,\ell}\right),
\label{channel_matrix}
\end{equation}
where $g_{k,\ell}$ represents the small-scale fading gain of the $\ell$-th sub-path in the $k$-th cluster, and $\mathbf{a}_{\rm bs}(\cdot)\in\mathbb{C}^{N_{\rm bs}\times 1}$ and $\mathbf{a}_{\rm ue}(\cdot)\in\mathbb{C}^{N_{\rm ue}\times 1}$ denote the $3$-Dimensional (3D) responses of the BS and UE antenna arrays, respectively, for each sub-path of every channel cluster. The angles $\theta^{\rm bs}_{k,\ell}$ and $\phi^{\rm bs}_{k,\ell}$ refer respectively to the vertical and horizontal Angles of Arrival (AoA) of the $\ell$-th sub-path in the $k$-th cluster, while $\theta^{\rm ue}_{k,\ell}$ and $\phi^{\rm ue}_{k,\ell}$ are respectively the vertical and horizontal angles of departure of the $k$-th cluster's $\ell$-th sub-path. The sub-path gain $g_{k,\ell}$ is further modeled as
\begin{align}
g_{k,\ell}=\sqrt{P_{k,\ell}}\exp\left(-j2\pi \tau_{k,\ell}f_c\right),
\label{scale_fading}
\end{align}
where $P_{k,\ell}$ denotes the power gain of the $\ell$-th sub-path in the $k$-th cluster, $\tau_{k,\ell}$ is the delay spread induced by different sub-path distances, and $f_c$ indicates the carrier frequency.

\vspace{-0.5mm}
\section{Multi-Level Beam Codebook Design}\label{sec:Codebook_Design}
Recent beam codebook designs for mmWave HBF systems follow several driving principles \cite{Hosoya_TAP2015_all, Raghavan_JSTSP, Aykin_2020}. In general, analog BF is targeted to compensate for the high propagation losses. However, a drawback of highly directive transmissions is their intolerance to beam misalignment. Digital Fourier Transform (DFT) beams yield the highest array gain \cite{Hosoya_TAP2015_all}, but even an error of few degrees leads to reduced link budget. Different approaches are hence necessary to encompass all cases where the optimal AoA is unknown a priori, for instance during the beam training procedure. In this paper, we propose the adoption of FTBs, which are beams of an almost constant value for a large range of angles around the direction of transmission. 
\vspace{-0.5mm}
\subsection{Flat Top Beam (FTB) Generation}
Ideal FTBs are equivalent to rectangular pulses in the angular domain. Since the array angular response is obtained from the Fourier transform of the weight distribution of the array's elements, one can generate analog beams through the inverse DFT. However, with this approach, the resulting weight distribution follows a sinc function, which imposes amplitude control on the weights. This contrasts with the HBF hardware constraints for analog BF, which only allows setting the phase of the antenna elements or deactivating them, but does not allow to fine tune the amplitude of the weight of any element. 

\begin{figure}[!t]
	\centering
	\includegraphics[width=0.58\columnwidth]{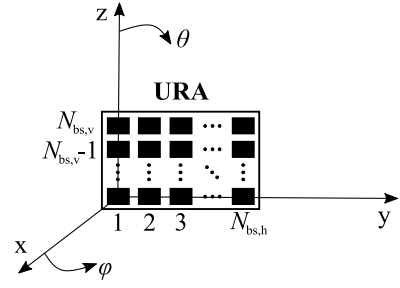}
	\caption{Geometrical model of the $N_{\rm bs,h}\times N_{\rm bs,v}$ URA at the BS side positioned on the $yz$ axis. The angles $\theta$ and $\phi$ represent the inclination and azimuth angles, respectively.}
	\label{Fig:GeomScenario}
	\vspace{-5mm}
\end{figure}

To overcome the constant amplitude limitation for the analog beams, we commence by noting that errors in a node's position estimation are more common in the azimuth rather than in inclination. We therefore impose the flat part of the beam to be only in the azimuth plain, while keeping the standard DFT beam shape for inclination. The basic idea of the proposed technique stems from the fact that the angular response of a rectangular array on the azimuth is equivalent to the one of a linear array, whose weights are the sum of the antenna elements on the vertical axis of the rectangular array. Hence, it is possible to approximate the sine cardinal values resulting in a FTB only in the azimuth. More specifically, the array angular response for the case of isotropic radiators as antenna elements is known as the Array Factor (AF) and is defined for the $N_{\rm bs}$-element antenna array at the BS as \cite{AntennaArray}

\begin{equation}
\label{eq:arrayfactor}
{\rm AF}\left(\theta,\phi \right) = \sum_{n=1}^{N_{\rm bs}} w_n \exp\left(-j \mathbf{k}^{\rm{H}}\mathbf{r}_n \right),
\end{equation}
where $j\triangleq\sqrt{-1}$ and $\theta,\phi\in[0,2\pi]$ are the inclination and azimuth angles, respectively, while $\mathbf{k}\in\mathbb{R}^{3\times1}$ and $\mathbf{r}_n\in\mathbb{R}^{3\times1}$ are respectively the wavevector and the 3D position of the $n$-th ($n=1,2,\ldots,N_{\rm bs}$) radiating antenna element. AF captures an array's angular response as a function of its geometrical properties. It is possible to state the following equivalence between the AF of a Uniform Rectangular Array (URA), ${\rm AF}_{\rm R}(\theta,\phi)$, with weights $[\mathbf{w}_{\rm R}]_{n,m}$ where $n= 0,1,\ldots,N_{\rm bs,h}-1$ and $m=0,1,\ldots,N_{\rm bs,v}-1$ and the AF of a Uniform Linear Array (ULA), ${\rm AF}_{\rm L}(\theta,\phi)$, with weights $[\mathbf{w}_{\rm L}]_{n}$.
\begin{proposition}\label{TH:proposition}
For the case $[\mathbf{w}_{\rm L}]_{n} = \sum_{m=0}^{N_{\rm bs,v}-1} [\mathbf{w}_{\rm R}]_{n,m}$, it holds $\forall$$\phi\in[0,2\pi]$ that ${\rm AF}_{\rm R}(\pi/2,\phi) = {\rm AF}_{\rm L}(\pi/2,\phi)$.
\end{proposition}
\begin{proof}
Assume that the $N_{\rm bs,h}\times N_{\rm bs,v}$ URA is placed on the $zy$-plane, as illustrated in Fig.\ref{Fig:GeomScenario}. Its corresponding AF can be written using the definition in \eqref{eq:arrayfactor} as ($d$ and $\lambda$ denote the antenna spacing and signal wavelength, respectively):
\begin{equation}\label{EQ:AF_RECT}
\begin{split}
&{\rm AF}_{\rm R}\left(\theta,\phi\right) = \sum_{m=0}^{N_{\rm bs,v}-1}\sum_{n=0}^{N_{\rm bs,h}-1}[\mathbf{w}_{\rm R}]_{n,m}
\\&\times\exp{\left( j\frac{2\pi d}{\lambda}(m\cos(\theta)+n\sin(\theta)\cos(\phi))\right)}.
\end{split}
\end{equation}
We also assume a $N_{\rm bs,h}$-element ULA disposed on the $y$-axis with the following AF:
\begin{equation}
\label{EQ:AF_LIN}
{\rm AF}_{\rm L}\left(\theta,\phi \right) = \sum_{n=0}^{N_{\rm bs,h}-1}[\mathbf{w}_{\rm L}]_{n}\exp{\left( j\frac{2\pi d}{\lambda}n\sin(\theta)\cos(\phi)\right)}.
\end{equation}
Evaluating \eqref{EQ:AF_RECT} at the horizon (i$.$e$.$, for $\theta =\pi/2$) and using the definition $[\mathbf{w}_{\rm L}]_{n} = \sum_{m=0}^{N_{\rm bs,v}-1} [\mathbf{w}_{\rm R}]_{n,m}$, it is possible to write the expression: 
\begin{equation}\label{EQ:AF_LINEAR}
\begin{aligned}
{\rm AF}_{\rm R}\left(\frac{\pi}{2},\phi\right) & =  \sum_{m=0}^{N_{\rm bs,v}-1}\sum_{n=0}^{N_{\rm bs,h}-1}[\mathbf{w}_{\rm R}]_{n,m}\exp{\left( j\frac{2\pi d}{\lambda}n\cos(\phi) \right)}  \\
 & = \sum_{n=0}^{N_{\rm bs,h}-1}\exp{\left( j\frac{2\pi d}{\lambda}n\cos(\phi) \right)}\sum_{m=0}^{N_{\rm bs,v}-1}[\mathbf{w}_{\rm R}]_{n,m}  \\
 & =  \sum_{n=0}^{N_{\rm bs,h}-1}[\mathbf{w}_{\rm R}]_{n}\exp{\left( j\frac{2\pi d}{\lambda} n\cos(\phi)\right)}.
\end{aligned}
\end{equation}
Finally, substituting $\theta =\pi/2$ into \eqref{EQ:AF_LIN} yields \eqref{EQ:AF_LINEAR}'s right-hand side, which completes the equivalence proof. 
\end{proof}

Proposition~\ref{TH:proposition} indicates that the problem of designing particular angular responses for $N_{\rm bs,h}\times N_{\rm bs,v}$ URAs with constant amplitude weights can be transformed into the design of the same angular response with $N_{\rm bs,h}$-element ULAs having weights with tunable amplitude and phase values. We hence refer to $[\mathbf{w}_{\rm L}]_{n}$ as the \textit{linear equivalent} of the $[\mathbf{w}_{\rm R}]_{n,m}$ URA weight. The design of the URA weights for a FTB can hence follow the two-step procedure: \textit{i}) design a FTB for a ULA by sampling an appropriate sinc; and \textit{ii}) approximate the derived samples with the weights of the URA such that $\sum_{m=0}^{N_{\rm bs,v}-1} [\mathbf{w}_{\rm R}]_{n,m} \cong [\mathbf{w}_{\rm L}]_{n}$, where $[\mathbf{w}_{\rm L}]_{n}$ is obtained as \begin{equation}\label{EQ:wneq}
[\mathbf{w}_{\rm L}]_{n}\!\!=\!\!\left\lbrace
\begin{array}{ll}
N_{\rm bs,h} \sinc \left(\frac{n-0.5(N_{\rm bs,v}-1)}{a} \right), &\!\!\rm{if}\,\,N_{\rm bs,v}\,\,\rm{even} \\
N_{\rm bs,h}  \sinc \left(\frac{n-0.5N_{\rm bs,v}}{a} \right), &\!\!\rm{if}\,\,N_{\rm bs,v}\,\,\rm{odd}
\end{array}\right.\!\!\!\!.
\end{equation}
In the latter expression, $a\in\mathbb{R}$ is a tuning parameter for the width of the sinc controlling the FTB's width. Finally, on each column of the URA, a number of antenna elements is turned on until $\sum_{m=0}^{N_{\rm bs,v}-1} [\mathbf{w}_{\rm R}]_{n,m}\geq [\mathbf{w}_{\rm L}]_{n}$. This beam design procedure creates centered FTBs of variable width, whose steering vectors can be easily obtained by multiplying the resulting weights with a steering matrix.
\vspace{-0.5mm}
\subsection{$L$-Level Hierarchical Beam Codebook}
\begin{figure}[!t]
\centering
\includegraphics[width=0.48\textwidth]{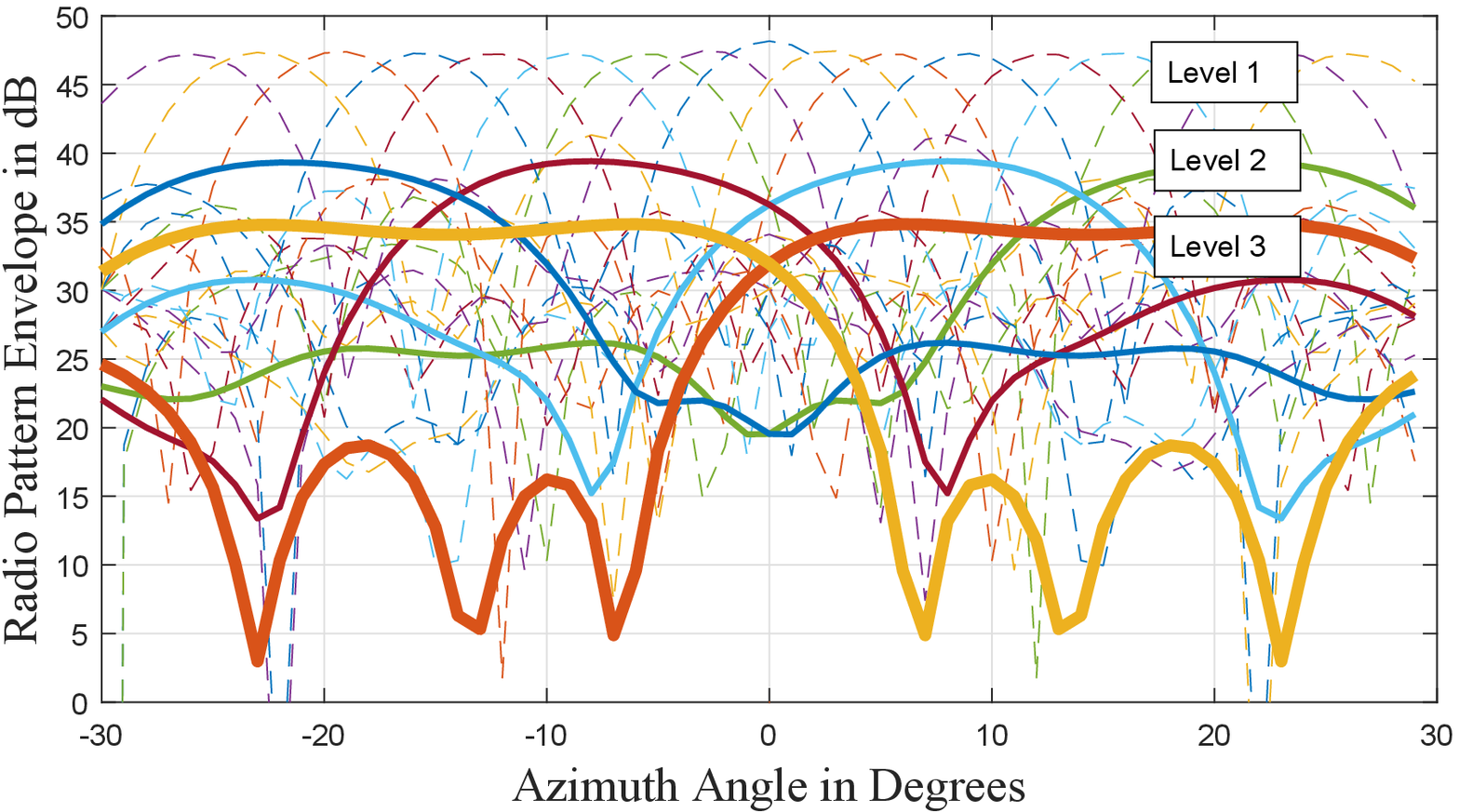}
\caption{Three-level beam codebook for a $16\times16$ URA comprising of $K_1=17$ narrow DFT beams in the first level, $K_2=4$ wide FTBs in the second level, and $K_3=2$ very wide FTBs in the final third level.}
\label{Fig:Three_Level_Codebook}
\vspace{-5mm}
\end{figure}

In this paper, we consider a hierarchical beam codebook with $L$ levels of different beam widths for each $i$-th BS AiP and use the notation $\varphi_k^{[\ell]}\in[-\pi,\pi]$ with $\ell=1,2,\ldots,L$ and $k=1,2,\ldots,K_\ell$ to represent the center angle of the $k$-th beam at the $\ell$-th level. The latter angles at each $\ell$-th level are assumed indexed in increasing values in the angle set $\mathbb{K}_\ell$; clearly, ${\rm card}(\mathbb{K}_\ell)=K_\ell$ and $\sum_{\ell=1}^LK_\ell={\rm card}(\mathbb{F}_{\rm bs})$. We also use hereinafter the notation $\mathbf{u}_i(\varphi_k^{[\ell]})$ to represent the BF vector realizing the beam in the $\ell$-th level centered at $\varphi_k^{[\ell]}$. Without loss of generality, we assume that the first level contains the beams in the hierarchical beam codebook with the narrowest widths (e.g., DFT beams), while the $L$-th level includes the FTBs with the largest widths. Beams resulting from angles of the lower levels are usually included into beams centered at upper levels' angles. We use notation $\mathbb{M}_{k,i}^{[\ell]}$ to represent the angle set in the $\ell$-th codebook covered by the $k$-th FTB of the $(\ell+1)$-th level codebook at $\varphi_k^{[\ell+1]}$. Figure~\ref{Fig:Three_Level_Codebook} depicts an example hierarchical beam codebook with $L=3$ including two levels of FTBs with different widths and steering angles for a $16\times16$ URA. As shown in the figure, each beam resulting from angles in $\mathbb{K}_3$ shares common coverage with approximately $2$ beams from angles in $\mathbb{K}_2$, which in turn have overlapping coverage with $4$ beams from angles in $\mathbb{K}_1$. For example, the weights for the second from the left FTB centered at the $\mathbb{K}_2$ angle $\varphi_2^{[2]}=-8 ^\circ$ were obtained using the following $16\times16$ weight matrix: 
\begin{equation}\label{Eq:PrecMatrix} \small
\mathbf{W}(\varphi_2^{[2]})\!=\!\left[
\begin{array}{cccccccccccccccc}
 &\mathbf{0}_{5\times3} &           	           &&                       \\
-1 &  1 & 1 &      		               &&                       \\
-1 & -1 & 1 & 				               &&                       \\  
-1 & -1 & 1 &     			             &&                       \\
-1 & -1 & 1 & \mathbf{0}_{16\times4} & -\mathbf{1}_{16\times4} & \mathbf{0}_{16\times5}\\
-1 & -1 & 1 &        		             &&                       \\
-1 & -1 & 1 &				                 &&                       \\
-1 &  0 & 1 &			                   &&                       \\
&  \mathbf{0}_{4\times3}  &     			             &&                       \
\end{array}
\right]\!\!,
\end{equation}
which highlights the link between a radiating element position and its corresponding weight. The analog BF vector for this URA beam can be obtained by simply stacking $\mathbf{W}(\varphi_2^{[2]})$'s elements into a column vector with $16^2=256$ elements. If this analog beam is deployed in all $M_{\rm bs}$ BS AiPs, then it suffices to set $\mathbf{u}_i(\varphi_2^{[2]})={\rm vec}(\mathbf{W}(\varphi_2^{[2]}))$ in \eqref{EQ:PC_Analog_Combiner} $\forall$$i=1,2,\dots,M_{\rm bs}$. 

\vspace{-0.5mm}
\section{Beam Management for HBF Mobility}\label{sec:Algos}
In 5G NR, BM for the control of multiple beams at frequencies above $6\,{\rm GHz}$ is composed of four different procedures \cite{J:5G_NR_2017}: \textit{i}) \textit{beam sweeping} for covering a spatial area with a set of beams; \textit{ii}) \textit{beam measurement} for the evaluation of the quality of the received signal (e$.$g$.$, its Signal-to-Noise Ratio (SNR) value); \textit{iii}) \textit{beam determination} for the selection of the suitable beam pairs; and \textit{iv}) \textit{beam reporting} for communicating the BF quality and decision information to the radio access network. The latter procedures are periodically repeated to update the optimal BS and UE beam pairs over time. To monitor UL channel quality, the UE transmits multiple Sounding Reference Signals (SRSs) to the BS, and the BS notifies the UE which resource to use for UL transmission. The BS and UE roles are reversed when the Downlink (DL) channel quality needs to be estimated. According to the latest BM definition for the UL \cite{3GPP_NR_ArXiv2018}, SRSs can span up to $4$ orthogonal frequency division multiplexing symbols in each Transmission Time Interval (TTI). Each resource for channel sounding may be periodic (configured at the TTI level), semi-persistent (also at TTI level, but it can be activated or deactivated with BS messages), and aperiodic (sounding is triggered by the BS). In this paper, we consider the periodic case.
\vspace{-0.5mm}
\subsection{Proposed Beam Management Scheme}\label{sec:Proposed_Beam_Training}
By considering reciprocal UL and DL communications, we assume that at the first TTI (i.e., upon BM initialization), the beam searching commences in the DL direction using the $K_1$ narrow beams of the first level for each BS AiP. Specifically, at the first CTS of this TTI, all $M_{\rm bs}$ AiPs transmit simultaneously orthogonal pilot symbols (i.e., SRSs) to the UE using a narrow beam centered at $\varphi_1^{[1]}$; in particular, each $i$-th BS AiP beamforms in the analog domain its pilot symbols with $\mathbf{u}_i(\varphi_1^{[1]})$. Consequently, the UE measures the SNR at each of its $N_{\rm ue}$ antenna branches for the pilot symbols received by each BS AiP. Then, it computes the summation of the SNR values from all UE antennas referring to each $\mathbf{u}_i(\varphi_1^{[1]})$ of the $i$-th BS AiP. We term each of the latter summations as $\gamma_i(\varphi_1^{[1]})$, which actually represents the output SNR with Maximal Ratio Combining (MRC) at the UE. Finally, the quantity $\sum_{i=1}^{M_{\rm bs}}\gamma_i(\varphi_1^{[1]})$ is compared to the SNR threshold $\gamma_{{\rm th}}$, which refers to the link's Quality-of-Service (QoS) requirement. In general, $\gamma_{{\rm th}}$ may change with the evolution of TTIs, implying time varying QoS requirements; this can be straightforwardly incorporated to the proposed approach.
If $\gamma_{{\rm th}}$ is met with the latter AiPs' configuration, beam alignment is achieved at this first CTS of the first TTI, and the UE notifies the BS to realize $\mathbf{u}_i(\varphi_1^{[1]})$ at each $i$-th BS AiP for the upcoming UL data transmission. For this data communication, the UE first estimates the beamformed DL MIMO channel $\mathbf{H}^{\rm H}\mathbf{U}_{\rm RF}\in\mathbb{C}^{N_{\rm ue}\times M_{\rm bs}}$ from pilot symbols and then applies Maximal Ratio Transmission (MRT). However, if $\gamma_{\rm th}$ is not satisfied with $\mathbf{u}_i(\varphi_1^{[1]})$'s, the remaining $K_1-1$ beams of $\mathbb{K}_1$ are searched sequentially until an acceptable AiPs' configuration is identified, or the maximum allowable number of CTSs is reached. During the latter repetitive procedure, the UE is considered capable of storing $\gamma_i(\varphi_j^{[1]})$ for each $i$-th BS AiP and for each $j$-th realized angle (with $j=1,2,\ldots$) within each TTI. It hence considers, at each $j$-th CTS, the AiPs' configuration yielding the maximum received SNR up to this slot in the comparison with the SNR threshold $\gamma_{{\rm th}}$. 

The previously described BM initialization resulted in the BS AiPs' configuration $\mathbf{u}_i(\varphi_{\nu_{i,1}}^{[1]})$ for each $i$-th BS AiP, where notation $\nu_{i,\ell}$ represents the angle index in the $\ell$-th level angle set $\mathbb{K}_\ell$. This analog combining is used till an UL beam misalignment event is identified, which means that the QoS requirement is not met. Suppose that this happens at the $(t+1)$-th TTI triggering the proposed multi-level BM approach summarized in Algorithm~\ref{trial} for $L=3$, which incorporates an UL phase with narrow beam sweeping followed by a DL phase with $3$-level hierarchical beam searching, if necessary. It is noted that the latter phase can be trivially modified to $L$-level beam searching with $L=2$ or $L>3$. As per the 5G NR limitations \cite{3GPP_NR_ArXiv2018}, UL beam sweeping spans the first $4$ CTSs of each $(t+1)$-th TTI, during which the BS searches over $4$ analog combining vectors in the first level beam codebook for all its AiPs. The sequence of the $4$ angles to be swept for the latter vectors is designed using Algorithm~\ref{Beam_Sweeping}, which targets analog combining searching over adjacent angles to those decided for the previous $t$-th TTI. In particular, for each $i$-th AiP, this algorithm computes the angle sweeping sequence vector $\mathbf{m}_i^{[\ell]}$ having ${\rm card}\left(\mathbb{K}_\ell\right)-1$ elements referring to the angles' indices; the angle index used in the previous $t$-th TTI is excluded. 
\begin{algorithm}[t!]\caption{$3$-Level BM at the $(t+1)$-th TTI}\label{trial}
\begin{algorithmic}[1]
\State \textbf{Input:} $\mathbb{K}_\ell$, $K_\ell$, and $\nu_{i,\ell}$ for the $t$-th TTI $\forall$$i=1,2,\ldots,M_{\rm bs}$ and for $\ell=1,2$, and $3$, $\mathbb{F}_{\rm bs}$, and $\mathbf{v}_{\rm BB}^{({\rm MRT})}$ of the
latest UL data communication. 

\noindent\textbf{\textit{UL Beam Management Phase:}}
\State Find the angle sweeping sequence $\mathbf{m}_i$ for the first level beam codebook for this $(t+1)$-th TTI given $\nu_{i,1}$ for the $t$-th TTI for each $i$-th BS AiP using Algorithm~\ref{Beam_Sweeping}.
\For {$j=1,2,3$, and $4$}
    \State During the $j$-th CTS, compute $\gamma(\varphi_{[\mathbf{m}_i]_j}^{[1]})$ for each 		
		
		\hspace{-0.18cm}$i$-th BS AiP using \eqref{Eq:UL_Received_Signal_SNR}.
\EndFor {}		
    \State Compute $\gamma_i$ for each $i$-th BS AiP using \eqref{Eq:max_per_AiP}.
\If {$\sum_{i=1}^{M_{\rm bs}}\gamma_i\geq\gamma_{\rm th}$},
    \State Obtain $\varphi_{[\mathbf{m}_i]_{\rm max}}^{[1]}$ for each $i$-th BS AiP.
		\State \textbf{Output:} $\mathbf{u}_i(\varphi_{[\mathbf{m}_i]_{\rm max}}^{[1]})$ for each $i$-th BS AiP.
\EndIf {}

\noindent\textbf{\textit{DL Beam Management Phase:}}
\State For each $i$-th BS AiP, exclude the first level beams swept in the UL phase from each $r$-th angle set $\mathbb{M}^{[1]}_{r,i}$ with $r=1,2,\ldots,K_2$ to obtain the new sets $\bar{\mathbb{M}}^{[1]}_{r,i}$.
\State Run Algorithm~\ref{Beam_Sweeping} to obtain $\mathbf{m}_i^{[3]}$ for the third level beam codebook of each $i$-th BS AiP using as $\nu_{i,3}$ the index of the angle in $\mathbb{K}_3$ resulting in the FTB $\mathbf{u}_i(\varphi_{\nu_{i,3}}^{[3]})$ that includes $\varphi_{[\mathbf{m}_i]_{\rm max}}^{[1]}$ of Step $8$. Then, set the angle sweeping sequence for each of those beam codebooks as $\mathbf{p}_i^{[3]}=[\nu_{i,3}\,\,\mathbf{m}_i^{[3]}]$.
\State For each of the angle sets $\bar{\mathbb{M}}^{[1]}_{r,i}$ of Step $11$ and $\mathbb{M}^{[2]}_{k,i}$ with $k=1,2,\ldots,K_3$, obtain their angle sweeping sequences $\mathbf{p}_{r,i}^{[1]}$ and $\mathbf{p}_{k,i}^{[2]}$, respectively, similar to Steps $2$ and $12$.

\For {$k=1,2\ldots,\text{length}(\mathbf{p}_i^{[3]})$}
\For {$m=1,2\ldots,\text{length}(\mathbf{p}_{k,i}^{[2]})$}
\For {$n=1,2\ldots,\text{length}(\mathbf{p}_{m,i}^{[1]})$}	
    \State During the $(\rho+4)$-th CTS, compute $\gamma(\bar{\varphi}_{n,i}^{[1]})$ 
		
		\hspace{0.89cm}for each $i$-th BS AiP similar to \eqref{Eq:UL_Received_Signal_SNR}, where
		
		\hspace{0.89cm}$\gamma(\bar{\varphi}_{n,i}^{[1]})$ represents the angle $\bar{\mathbb{M}}^{[1]}_{m,i}\{[\mathbf{p}_{m,i}^{[1]}]_n\}$.
\If {$\sum_{i=1}^{M_{\rm bs}}\gamma(\bar{\varphi}_{n,i}^{[1]})\geq\gamma_{\rm th}$},
		\State \textbf{Output:} $\mathbf{u}_i(\bar{\varphi}_{n,i}^{[1]})$ for each $i$-th BS AiP.
\EndIf {}
\State Set $\rho=\rho+1$.
\EndFor{}
\EndFor{}
\EndFor{}
\State \textbf{Output:} Beam alignment was not achieved in the current $(t+1)$-th TTI, use this TTI's first level beams for the $M_{\rm bs}$ BS AiPs resulted in the best performance for BM initialization at the $(t+2)$-th TTI. 
\end{algorithmic}
\end{algorithm}

In Step $4$ of Algorithm~\ref{trial}, the instantaneous received SNR at the end of each $j$-th CTS at each $i$-th BS AiP that realizes the analog BF vector $\mathbf{u}_i(\varphi_{[\mathbf{m}_i]_j}^{[1]})$ is calculated as
\begin{equation}\label{Eq:UL_Received_Signal_SNR}
\gamma(\varphi_{[\mathbf{m}_i]_j}^{[1]})\triangleq\frac{{\rm P}|\mathbf{u}_i^{\rm H}(\varphi_{[\mathbf{m}_i]_j}^{[1]})\mathbf{H}_i\mathbf{v}_{\rm BB}^{({\rm MRT})}|^2}{\sigma^2\|\mathbf{u}_i^{\rm H}(\varphi_{[\mathbf{m}_i]_j}^{[1]})\|^2}.
\end{equation} 
Next, at the end of each $4$-th CTS of each $(t+1)$-th TTI, the maximum received SNR for each $i$-th BS AiP is obtained as
\begin{equation}\label{Eq:max_per_AiP}
\gamma_i = \max\{\gamma(\varphi_{[\mathbf{m}_i]_1}^{[1]}),\gamma(\varphi_{[\mathbf{m}_i]_2}^{[1]}),\gamma(\varphi_{[\mathbf{m}_i]_3}^{[1]}),\gamma(\varphi_{[\mathbf{m}_i]_4}^{[1]})\},
\end{equation}
and the quantity $\sum_{i=1}^{M_{\rm bs}}\gamma_i$ is calculated and finally compared to the SNR threshold value $\gamma_{\rm th}$. If this performance threshold is met, beam alignment is achieved with the BS AiPs' configuration yielding $\gamma_i$'s; we denote the first level beam angle yielding $\gamma_i$ for the $i$-th BS AiP by $\varphi_{[\mathbf{m}_i]_{\rm max}}^{[1]}$. Otherwise, the DL BM phase of Algorithm~\ref{trial} is followed, which includes our novel $3$-level hierarchical beam searching.  
\begin{algorithm}[t!]\caption{Angle Sweeping Sequence at the $i$-th BS AiP}\label{Beam_Sweeping}
\begin{algorithmic}[1]
\State \textbf{Input:} ${\rm card}\left(\mathbb{K}_\ell\right)$, $\nu_{i,\ell}$, $\mu=1$, $\mathbf{m}_i^{[\ell]}=\mathbf{0}_{({\rm card}\left(\mathbb{K}_\ell\right)-1)\times1}$.
\If{$\nu_{i,\ell}+1\leq{\rm card}\left(\mathbb{K}_\ell\right)$,}
	\State $[\mathbf{m}_i^{[\ell]}]_1=\nu_{i,\ell}+1$.
\For{$p=3,4,\ldots,{\rm card}\left(\mathbb{K}_\ell\right)$}
  \State $\mu=-\mu - \min\left(0,\frac{\mu}{|\mu|}\right)$; $\xi=\nu_{i,\ell}+\mu$.
\If{$\xi>{\rm card}\left(\mathbb{K}_\ell\right)$ or $\xi\leq0$,}	
  \State $\mu=-\mu - \min\left(0,\frac{\mu}{|\mu|}\right)$; $[\mathbf{m}_i^{[\ell]}]_{p-1}=\nu_{i,\ell}+\mu$. 
\Else
  \State $[\mathbf{m}_i^{[\ell]}]_{p-1}=\xi$. 
\EndIf			
\EndFor	
\Else
   \State $[\mathbf{m}_i^{[\ell]}]_1={\rm card}\left(\mathbb{K}_\ell\right)-1$.
\For{$p=3,4,\ldots,{\rm card}\left(\mathbb{F}_{\rm bs}\right)$}
  \State $[\mathbf{m}_i^{[\ell]}]_{p-1}={\rm card}\left(\mathbb{K}_\ell\right)-i+1$.
\EndFor		
\EndIf	
\State \textbf{Output:} $\mathbf{m}_i^{[\ell]}$ including the beam sweeping sequence. 
\end{algorithmic}
\end{algorithm}

\vspace{-0.5mm}
\section{Numerical Results and Discussion}\label{sec:Results}
In this section, we evaluate the performance of the proposed BM algorithm using the $3$-level beam codebook of Fig$.$~\ref{Fig:Three_Level_Codebook}. Both the average number of required beam searches to meet a desired QoS threshold (which indicates the latency requirements for BM) and the average received SNR in baseband have been simulated over the channel model in Section~\ref{Sec:Channel_Model}, using the parameters' setting of \cite{rebato19_all}: $K=4$ clusters each composed of $L=10$ subpaths. In the results that follow, apart from the proposed Algorithm~\ref{trial}, we have evaluated the performance of the exhaustive narrow beam search that leads to the highest received SNR. In addition, we have considered two variations of the proposed BM for comparison: \textit{i}) 'Only UL BM' that implements only the UL BM phase of Algorithm~\ref{trial}; and \textit{ii}) 'DL-Assisted UL BM' that is Algorithm~\ref{trial} with only first level beam sweeping at the DL BM phase. For all cases, the UE performs optimum spatial filtering via the eigendecomposition of the MIMO channel.
\begin{figure}[!t]
		\centering
        \includegraphics[width=0.43\textwidth]{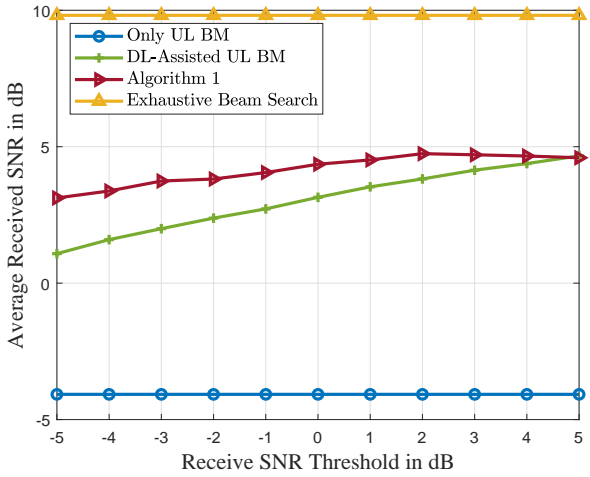}
        \caption{Average received SNR in dB versus the receive SNR threshold in dB for transmit SNR set to $-1\,{\rm dB}$.}
				\label{fig:TXminus1dB}\vspace{-5mm}
\end{figure}

\begin{figure}[!t]
    \centering
        \includegraphics[width=0.43\textwidth]{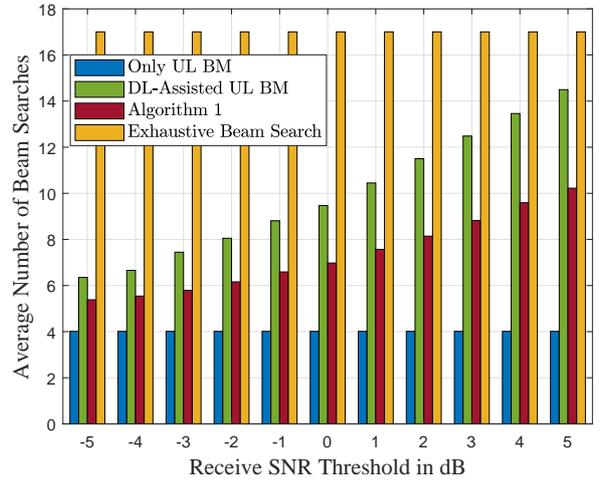}
				\caption{Average number of beam searches for the setting of parameters in Fig$.$~\ref{fig:TXminus1dB}.}
				\label{fig:beam_searches_TXminus1dB}
    \end{figure}

\begin{figure}[!t]
	\centering
			\includegraphics[width=0.43\textwidth]{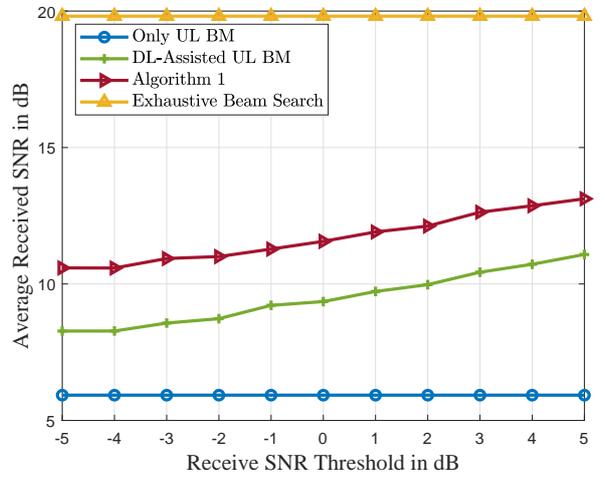}
			\caption{Average received SNR in dB versus the receive SNR threshold in dB for transmit SNR set to $0\,{\rm dB}$.}
			\label{fig:TX0dB}\vspace{-5mm}
	\end{figure}

\begin{figure}[!t]
\centering
		\includegraphics[width=0.43\textwidth]{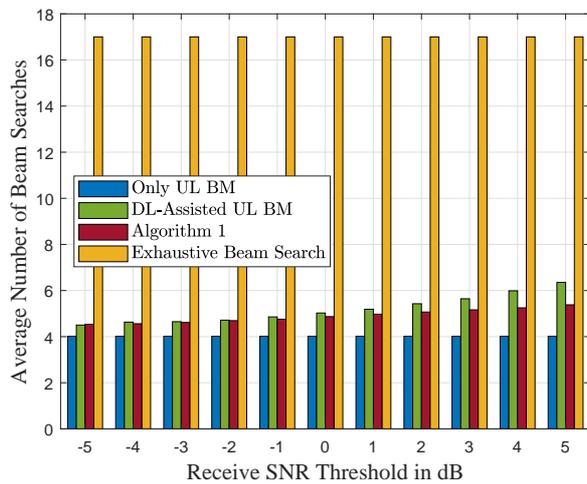}
		\caption{Average number of beam searches for the setting of parameters in Fig$.$~\ref{fig:TX0dB}.}
		\label{fig:beam_searches_TX0dB}\vspace{-5mm}
\end{figure}

The average received SNR in dB as a function of the receive SNR threshold $\gamma_{{\rm th}}$ in dB for the transmit SNR value $-1\, {\rm dB}$ (defined as ${\rm P}/\sigma^2$) is illustrated in Fig$.$~\ref{fig:TXminus1dB}, whereas Fig$.$~\ref{fig:beam_searches_TXminus1dB} includes the respective average number of beam searches. In Fig$.$~\ref{fig:TX0dB} and~\ref{fig:beam_searches_TX0dB}, we plot similar results for the average received SNR and number of beam searches, respectively, considering that the transmit SNR has increased to $0\, {\rm dB}$. As expected, a larger $\gamma_{{\rm th}}$ results in more beam searches for all compared BM schemes. The dependance of this metric on the transmit SNR can be understood by comparing the different trends in Figs$.$~\ref{fig:beam_searches_TXminus1dB} and~\ref{fig:beam_searches_TX0dB}. Higher transmit SNR results in significantly lower average number of beam searches. It is also shown that, for the same $\gamma_{{\rm th}}$, a higher transmit SNR results in improved performance for all considered schemes. In more detail, the exhaustive narrow beam search has a constant number of beam searches, exhausting all the Level $1$ available beams to achieve the highest received SNR; 
this scheme constitutes the upper bound for this metric. The 'Only UL BM' scheme uses the constant number of beam searches allowed by the UL part of Algorithm~\ref{trial}. This results in substantially lower average received SNR than the other three schemes, and in the inadequacy to achieve $\gamma_{{\rm th}}$ with low transmit SNR values. It is finally shown that the proposed Algorithm~\ref{trial}, which realizes hierarchical beam searching with the designed $3$-level beam codebook of Fig$.$~\ref{Fig:Three_Level_Codebook}, always meets $\gamma_{{\rm th}}$, performing closer to the exhaustive search case, and with the smallest average number of beam searches. 

\section{Conclusion and Future Work}
In this paper, we considered HBF transceivers with planar antenna panels and presented a multi-level beam codebook for the analog beamformer comprising FTBs with variable widths. Differently to the vast majority of the available research works, we focused on the UL communication and presented a novel BM algorithm with dynamic beam ordering. Our representative simulation results over a realistic mmWave channel model showcased that the proposed BM scheme provides latency performance improvement over various benchmark schemes, thus rendering it suitable for the stringent latency requirements of the latest mmWave standard discussions. We intend to extend the proposed scheme to multiple beams' alignment with HBF at both communication ends, and to point to multi-point communications. We also intend to design hierarchical phase profile codebooks for passive \cite{George_RIS_TWC2019_all} and active \cite{DMA2020} reconfigurable metasurfaces, and to devise low-latency phase profile configuration schemes.


\bibliographystyle{IEEEtran}
\bibliography{IEEEabrv,refs}

\begin{thebibliography}{10}
\providecommand{\url}[1]{#1}
\csname url@samestyle\endcsname
\providecommand{\newblock}{\relax}
\providecommand{\bibinfo}[2]{#2}
\providecommand{\BIBentrySTDinterwordspacing}{\spaceskip=0pt\relax}
\providecommand{\BIBentryALTinterwordstretchfactor}{4}
\providecommand{\BIBentryALTinterwordspacing}{\spaceskip=\fontdimen2\font plus
\BIBentryALTinterwordstretchfactor\fontdimen3\font minus
  \fontdimen4\font\relax}
\providecommand{\BIBforeignlanguage}[2]{{%
\expandafter\ifx\csname l@#1\endcsname\relax
\typeout{** WARNING: IEEEtran.bst: No hyphenation pattern has been}%
\typeout{** loaded for the language `#1'. Using the pattern for}%
\typeout{** the default language instead.}%
\else
\language=\csname l@#1\endcsname
\fi
#2}}
\providecommand{\BIBdecl}{\relax}
\BIBdecl

\bibitem{Rappaport_Access_all}
T.~S. Rappaport, S.~Sun, R.~Mayzus, H.~Zhao, Y.~Azar, K.~Wang, G.~Wong, J.~K.
  Schulz, M.~Samim, and F.~Gutierrez, ``Millimeter wave mobile communications
  for {5G} cellular: {I}t will work!'' \emph{{IEEE} {A}ccess}, vol.~1, pp.
  335--349, May 2013.

\bibitem{Akyildiz_COMmag2015}
I.~F. Akyildiz, C.~Han, and S.~Nie, ``Combating the distance problem in the
  millimeter wave and terahertz frequency bands,'' \emph{{IEEE} {C}ommun.
  {M}ag.}, vol.~55, no.~6, pp. 102--108, Jun. 2018.

\bibitem{J:IEEE_11ad}
``{IEEE} wireless {LAN MAC} and {PHY} specifications-- {A}mendment 3:
  {E}nhancements for very high throughput in the 60 {GH}z band,'' \emph{IEEE
  Std. 802.11ad}, 2012.

\bibitem{J:5G_NR_2017}
3GPP, ``{Study on New Radio (NR) Access Technology- Physical Layer Aspects-
  Release 14},'' TR 38.802, 2017.

\bibitem{Venkateswaran_ABF}
V.~Venkateswaran and A.-J. van~der Veen, ``Analog beamforming in {MIMO}
  communications with phase shift networks and online channel estimation,''
  \emph{{IEEE} {Trans}. {S}ignal {P}rocess.}, vol.~58, no.~8, pp. 4131--4143,
  Aug. 2010.

\bibitem{J:Doan_CMOS}
C.~H. Doan, S.~Emami, D.~A. Sobel, A.~M. Niknejad, and R.~W. Brodersen,
  ``Design considerations for 60 {GH}z {CMOS} radios,'' \emph{IEEE Commun.
  Mag.}, vol.~42, no.~12, pp. 132--140, Dec. 2004.

\bibitem{J:Molisch_HBF}
X.~Zhang, A.~F. Molisch, and S.-Y. Kung, ``Variable-phase-shift-based
  {RF}-baseband codesign for {MIMO} antenna selection,'' \emph{IEEE Trans.
  Signal Process.}, vol.~53, no.~11, pp. 4091--4103, Nov. 2005.

\bibitem{Molisch_HBF_2017_all}
A.~F. Molisch, V.~V. Ratnam, S.~Han, Z.~Li, S.~L.~H. Nguyen, L.~Li, and
  K.~Haneda, ``Hybrid beamforming for massive {MIMO}: {A} survey,''
  \emph{{IEEE} {Commun.} {M}ag.}, vol.~55, no.~9, pp. 134--141, Sep. 2017.

\bibitem{Alkhateeb_JSTSP_all}
A.~Alkhateeb, O.~E. Ayach, G.~Leus, and R.~W. Heath, Jr., ``Channel estimation
  and hybrid precoding for millimeter wave cellular systems,'' \emph{{IEEE}
  {J.} {S}el. {T}opics {S}ignal {P}rocess.}, vol.~8, no.~5, pp. 831--846, Oct.
  2014.

\bibitem{Marzi_TSP2016_all}
Z.~Marzi, D.~Ramasamy, and U.~Madhow, ``Compressive channel estimation and
  tracking for large arrays in mm-wave picocells,'' \emph{{IEEE} {T}rans.
  {S}ignal {P}rocess.}, vol.~11, no.~3, pp. 514--527, Apr. 2016.

\bibitem{Bazzi_SPAWC2016_all}
A.~Bazzi, D.~T.~M. Slock, L.~Meilhac, and S.~Panneerselvan, ``A comparative
  study of sparse recovery and compressed sensing algorithms with application
  to {AoA} estimation,'' in \emph{Proc. IEEE SPAWC}, Edinburgh, UK, 3-6 Jul.
  2016, pp. 1--5.

\bibitem{Alexandropoulos_GC16}
G.~C. Alexandropoulos and S.~Chouvardas, ``Low complexity channel estimation
  for millimeter wave systems with hybrid {A/D} antenna processing,'' in
  \emph{Proc. IEEE GLOBECOM}, Washington D.C., USA, 4-8 Dec. 2016, pp. 1--6.

\bibitem{Vlachos_SPL18}
E.~Vlachos, G.~C. Alexandropoulos, and J.~Thompson, ``Massive {MIMO} channel
  estimation for millimeter wave systems via matrix completion,'' \emph{{IEEE}
  {Si}gnal {P}rocess. {L}ett.}, vol.~25, no.~11, pp. 1675--1679, Nov. 2018.

\bibitem{Vlachos2019WidebandSampling}
------, ``{Wideband MIMO channel estimation for hybrid beamforming millimeter
  wave systems via random spatial sampling},'' \emph{IEEE J. Sel. Topics Signal
  Process.}, vol.~13, no.~5, pp. 1136--1150, Sep. 2019.

\bibitem{Zhang_WCNC2010_all}
H.~Zhang, S.~Venkateswaran, and U.~Madhow, ``Channel modeling and {MIMO}
  capacity for outdoor millimeter wave links,'' in \emph{Proc. IEEE WCNC},
  Sydney, Australia, 18-21 Apr. 2010, pp. 1--6.

\bibitem{Hur_TCOM_all}
S.~Hur, T.~Kim, D.~J. Love, J.~V. Krogmeier, T.~A. Thomas, and A.~Ghosh,
  ``Millimeter wave beamforming for wireless backhaul and access in small cell
  networks,'' \emph{{IEEE} {T}rans. {C}ommun.}, vol.~61, no.~10, pp.
  4391--4403, Oct. 2013.

\bibitem{Hosoya_TAP2015_all}
K.~Hosoya, N.~Prasad, K.~Ramachandran, N.~Orihashi, S.~Kishimoto,
  S.~Rangarajan, and K.~Maruhashi, ``Multiple sector {ID} capture ({MIDC}): A
  novel beamforming technique for 60 {GH}z band multi-{G}bps {WLAN/PAN}
  systems,'' \emph{{IEEE} {T}rans. {A}ntennas {P}ropag.}, vol.~63, no.~1, pp.
  81--96, Jan. 2015.

\bibitem{Wang_JSAC_all}
J.~Wang, Z.~Lan, C.-W. Pyo, T.~Baykas, C.-S. Sum, M.~A. Rahman, J.~Gao,
  R.~Funada, F.~Kojima, H.~Harada, and S.~Kato, ``Beam codebook based
  beamforming protocol for multi-{G}bps millimeter-wave {WPAN} systems,''
  \emph{{IEEE} {J}. {S}el. {A}reas {C}ommun.}, vol.~27, no.~8, pp. 1390--1399,
  Oct. 2009.

\bibitem{Jeong_COMmag2015}
C.~Jeong, J.~Park, and H.~Yu, ``Random access in millimeter-wave beamforming
  cellular networks: {I}ssues and approaches,'' \emph{{IEEE} {C}ommun. {M}ag.},
  vol.~53, no.~1, pp. 180--185, Jan. 2015.

\bibitem{Barati_2015}
C.~Barati, S.~A. Hosseini, S.~Rangan, P.~Liu, T.~Korakis, S.~S. Panwar, and
  T.~S. Rappaport, ``Directional cell discovery in millimeter wave cellular
  networks,'' \emph{{IEEE} {T}rans. {W}ireless {C}ommun.}, vol.~14, no.~12, pp.
  6664--6678, Dec. 2015.

\bibitem{Raghavan_JSTSP}
V.~Raghavan, J.~Cezanne, S.~Subramanian, A.~Sampath, and O.~Koymen,
  ``Beamforming tradeoffs for initial {UE} discovery in millimeter-wave {MIMO}
  systems,'' \emph{{IEEE} {J.} {S}el. {T}opics {S}ignal {P}rocess.}, vol.~10,
  no.~3, pp. 543--559, Apr. 2016.

\bibitem{Alexandropoulos_position}
G.~C. Alexandropoulos, ``Position aided beam alignment for millimeter wave
  backhaul systems with large phased arrays,'' in \emph{Proc. IEEE CAMSAP},
  Cura\c{c}ao, Dutch Antilles, 10-13 Dec. 2017, pp. 1--5.

\bibitem{Aykin_2020}
I.~Aykin and M.~Krunz, ``Efficient beam sweeping algorithms and initial access
  protocols for millimeter-wave networks,'' \emph{{IEEE} {T}rans. {W}ireless
  {C}ommun.}, vol.~19, no.~4, pp. 2504--2514, Apr. 2020.

\bibitem{Love_HBF_Codebook_2018}
J.~Song, J.~Choi, and D.~J. Love, ``Common codebook millimeter wave beam
  design: {D}esigning beams for both sounding and communication with uniform
  planar arrays,'' \emph{{IEEE} {T}rans. {W}ireless {C}ommun.}, vol.~65, no.~4,
  pp. 1859--1872, Apr. 2017.

\bibitem{akdeniz14_all}
M.~Akdeniz, Y.~Liu, M.~Samimi, S.~Sun, S.~Rangan, T.~Rappaport, and E.~Erkip,
  ``Millimeter wave channel modeling and cellular capacity evaluation,''
  \emph{IEEE J. Sel. Areas Commun}, vol.~32, no.~6, pp. 1164--1179, June 2014.

\bibitem{AiP_2009}
Y.~P. Zhang and D.~Liu, ``Antenna-on-chip and antenna-in-package solutions to
  highly integrated millimeter-wave devices for wireless communications,''
  \emph{{IEEE} {T}rans. {A}ntennas {P}ropag.}, vol.~57, no.~10, pp. 2830--2841,
  Oct. 2009.

\bibitem{winner2}
P.~Kyosti \emph{et~al.}, ``{WINNER} {II} channel model,'' \emph{Technical
  Report IST-WINNER D1.1.2 ver 1.1}, Sept. 2007.

\bibitem{AntennaArray}
R.~J. Mailloux, \emph{Phased Array Antenna Handbook}, New York, 2004.

\bibitem{3GPP_NR_ArXiv2018}
M.~{Giordani}, M.~Polese, A.~Roy, D.~Castor, and M.~Zorzi, ``{A tutorial on
  beam management for 3GPP NR at mmWave frequencies},'' \emph{IEEE Commun.
  Surveys Tuts}, vol.~21, no.~1, pp. 173--196, 2019.

\bibitem{rebato19_all}
M.~{Rebato}, L.~{Rose}, and M.~{Zorzi}, ``Performance assessment of {MIMO}
  precoding on realistic mm{W}ave channels,'' in \emph{Proc. IEEE ICC},
  Shanghai, China, 20-24 May 2019, pp. 1--6.

\bibitem{George_RIS_TWC2019_all}
C.~Huang, A.~Zappone, G.~C. Alexandropoulos, M.~Debbah, and C.~Yuen,
  ``{Reconfigurable intelligent surfaces for energy efficiency in wireless
  communication},'' \emph{IEEE Trans. Wireless Commun.}, vol.~18, no.~8, pp.
  4157--4170, Aug. 2019.

\bibitem{DMA2020}
N.~Shlezinger, G.~C. Alexandropoulos, M.~F. Imani, Y.~C. Eldar, and D.~R.
  Smith, ``Dynamic metasurface antennas for {6G} extreme massive {MIMO}
  communications,'' to appear, 2021, [online] https://arxiv.org/abs/2006.07838.

\end{thebibliography}
\end{document}